\documentclass[showkeys,showpacs]{revtex4}
\usepackage{amsmath}
\usepackage{amssymb}
\usepackage{graphicx}
\usepackage{color}
\usepackage{ulem}

\begin{document}

\title{
Modified gravity from the quantum part of the metric
}

\author{
Vladimir Dzhunushaliev,$^{1,2,3,4}$
\footnote{Email: v.dzhunushaliev@gmail.com}
Vladimir Folomeev,$^{2,3}$
\footnote{Email: vfolomeev@mail.ru}
Burkhard Kleihaus,$^{4}$
\footnote{Email: b.kleihaus@uni-oldenburg.de }
Jutta Kunz$^4$
\footnote{Email:  jutta.kunz@uni-oldenburg.de}
}
\affiliation{$^1$
Dept. Theor. and Nucl. Phys., KazNU, Almaty, 050040, Kazakhstan \\
$^2$ IETP, Al-Farabi Kazakh National University, Almaty, 050040, Kazakhstan \\
$^3$Institute of Physicotechnical Problems and Material Science of the NAS
of the
Kyrgyz Republic, 265 a, Chui Street, Bishkek, 720071,  Kyrgyz Republic \\
$^4$Institut f\"ur Physik, Universit\"at Oldenburg, Postfach 2503
D-26111 Oldenburg, Germany
}

\begin{abstract}
It is shown that if a metric in quantum gravity can be decomposed as a sum of classical and quantum parts
then Einstein quantum gravity looks approximately  like  modified gravity with a nonminimal interaction between gravity and matter.
\end{abstract}


\maketitle

\section{Introduction}

We know two strongly nonlinear physical theories: quantum chromodynamics and gravity. The nonlinearity in both theories are different:
in quantum chromodynamics, the nonlinearity is connected with potential terms but in gravity the nonlinearity appears in kinetic terms. The quantization of both theories has severe problems, because we only have quantization techniques for weakly interacting fields. In the 1950s, Heisenberg \cite{heis} has investigated a nonlinear spinor field theory and worked out a nonperturbative technique for such a kind of quantization. The essence of his method is to write an infinite set of equations for all Green's functions. Such a set of equations probably cannot be solved analytically and a way to solve such a set is to cut it off to obtain a finite set, taking into account some physical arguments for the cutoff. This situation is similar to turbulence modeling, where there exists a similar set of equations for all cumulants (see Ref.~\citep{Wilcox} for details).

The idea presented here is as follows: we consider a physical system in quantum gravity where the quantum metric $\mathcal G$ can be represented as a sum of a classical part, $g$, and a quantum part $\widehat g$. In doing so, we here assume that the expectation value of the quantum part, $\left\langle  \widehat g \right\rangle$, is not equal to zero. This gives rise to a modification of Einstein gravity, that results in the appearance of a nonminimal interaction between gravity and matter. To calculate the modified gravitational Lagrangian, we also assume  that the expectation value of the quantum part of the metric can be approximately presented as an expression  which depends on the classical part of the metric. Taking into account the aforementioned decomposition of the metric and retaining only terms up to first order in  $\left\langle  \widehat g \right\rangle$, we find the corresponding matter Lagrangian. As a result, we obtain modified gravity with a nonminimal interaction between gravity and matter.

Modified gravity theories are used both in modeling the evolution of the early Universe and in describing its present accelerated expansion.
In particular,
in recent years much success has been achieved in models where dark energy is described
on the basis of modified gravity (for a review, see Refs.~\cite{Nojiri:2010wj} and \cite{Bamba:2012cp}).

\section{Nonperturbative quantization technique}

When quantizing gravity,  there are several stages for performing
this procedure. In the first step, the metric is considered as an
ordinary (tensor) field. In the next step, one considers changes of topology, metric signature, etc.
Here we consider only the first stage of the quantizing procedure, using
Heisenberg's nonperturbative technique. According to this technique, one has
to use either the Einstein equations for operators of the metric and connections or (equivalently) an infinite set of equations for all Green's functions.

The operator Einstein equations are
\begin{equation}
	\hat R_{\mu \nu} - \frac{1}{2} \hat g_{\mu \nu} \hat R =
	\varkappa \hat T_{\mu \nu} ,
\label{1-10}
\end{equation}
where all geometrical quantities are defined in the usual manner from the corresponding operators
\begin{eqnarray}
	\hat R_{\mu \nu} &=& \hat R^\rho_{\phantom{\rho} \mu \rho \nu},
\label{1-20}\\
	\hat R^\rho_{\phantom{\rho} \sigma \mu \nu} &=&
	\frac{\partial \hat \Gamma^\rho_{\phantom{\rho} \sigma \nu}}
	{\partial x^\mu} -
	\frac{\partial \hat \Gamma^\rho_{\phantom{\rho} \sigma \mu}}
	{\partial x^\nu} +
	\hat \Gamma^\rho_{\phantom{\rho} \tau \mu}
	\hat \Gamma^\tau_{\phantom{\tau} \sigma \nu} -
	\hat \Gamma^\rho_{\phantom{\rho} \tau \nu}
	\hat \Gamma^\tau_{\phantom{\tau} \sigma \mu} .
\label{1-30}\\
	\hat \Gamma^\rho_{\phantom{\rho} \mu \nu} &=&
	\frac{1}{2} \hat g^{\rho \sigma} \left(
		\frac{\partial \hat g_{\mu \sigma}}{\partial x^\nu} +
		\frac{\partial \hat g_{\nu \sigma}}{\partial x^\mu} -
		\frac{\partial \hat g_{\mu \nu}}{\partial x^\sigma}
	\right).
\label{1-40}
\end{eqnarray}
Heisenberg's technique offers to use an infinite set of equations  for all Green's functions, which can be written as follows
\begin{eqnarray}
	\left\langle Q \left| \hat g (x_1)
	\cdot \text{ Eq. \eqref{1-10}}
	\right| Q \right\rangle &=& 0 ,
\label{1-50}\\
	\left\langle Q \left| \hat g(x_1) \hat g(x_2)
	\cdot \text{ Eq. \eqref{1-10} }
	\right| Q \right\rangle &=& 0 ,
\label{1-60}\\
	\cdots &=& 0	,
\label{1-150}\\
	\left\langle Q \left|
	\text{ the product of $g$ at different points $(x_1,
	\cdots , x_n)$} 	\cdot \text{ Eq. \eqref{1-10}}
	\right| Q \right\rangle &=& 0,
\label{1-70}
\end{eqnarray}
where $\left. \left|Q \right. \right\rangle$ is a quantum state (see Ref.~\cite{Dzhunushaliev:2012np} for details).

In all likelihood  the set of equations \eqref{1-50}-\eqref{1-70} cannot be solved analytically. But
there are two possibilities to find an approximate solution to these equations.
The first way consists in cutting off the
set of equations by using some decomposition of the form
 $G_{m+n} \approx G_m G_n$ (here $G_i$ is an $i$-point Green's function)
and taking into account only the first $p<m+n$ equations.
The second way is to take some functional (for example, an action) and
to average it using some assumptions about expectation values of  metric operators.

Here we  use the second way, decomposing the metric operator $\hat g_{\mu \nu}$ into  classical and quantum parts and 	
considering the expectation value of the quantum part as being non-zero. We then calculate the expectation value of the Lagrangian to
first order in
$\left\langle \delta g \right\rangle$. These calculations are performed in the manner similar to the ones used in
Ref.~\cite{Dzhunushaliev:2012vb} when considering  quantum torsion.

\section{Decomposition of the quantum metric}

The key idea we are dealing with here is that
the quantum metric $\mathcal G_{\mu \nu}$ can be decomposed into two parts: the classical metric $g_{\mu \nu}$ and the quantum metric
$\widehat g_{\mu \nu}$
\begin{equation}
  \mathcal G_{\mu \nu} = g_{\mu \nu} + \widehat g_{\mu \nu}
\label{2-10}
\end{equation}
together with the assumption that
\begin{equation}
  \left\langle \widehat g_{\mu \nu} \right\rangle \neq 0.
\label{2-20}
\end{equation}

In order to derive the modification of the Lagrangian induced by the quantum corrections $\left\langle \widehat g_{\mu \nu} \right\rangle$,
we have to expand the Lagrangian
to first order in
$\left\langle \widehat g_{\mu \nu} \right\rangle$.
To do this, we start from the Einstein-Hilbert Lagrangian
\begin{equation}
\mathcal L_{\mathcal G} = - \frac{c^2}{2 \varkappa} \sqrt{-\mathcal G} R,
\label{2-30}
\end{equation}
where $\varkappa=8\pi G/c^2$. Next,
to find corrections to this Lagrangian induced by the quantum corrections $\left\langle \widehat g_{\mu \nu} \right\rangle$,
we assume that the Lagrangian can be expanded in the following manner:
\begin{equation}
	\mathcal L_{\mathcal G}(g + \widehat{g}) \approx \mathcal L_g(g) +
	\frac{\delta \mathcal L_g}{\delta g^{\mu \nu}} \widehat{g}^{\mu \nu} =
  \mathcal L_g(g) + \sqrt{-g} \,G_{\mu \nu} \widehat{g}^{\mu \nu},
\label{2-40}
\end{equation}
where $G_{\mu \nu}$ is the Einstein tensor,
and
$$\mathcal L_g(g)= - \frac{c^2}{2 \varkappa} \sqrt{-g} R$$
denotes the classical (nonquantum) Lagrangian. In turn, the expectation value of
$\frac{\delta \mathcal L}{\delta g^{\mu \nu}} \widehat{g}^{\mu \nu}$ can be represented as follows:
\begin{equation}
 \left\langle
    \frac{\delta \mathcal L_g}{\delta g^{\mu \nu}} \widehat{g}^{\mu \nu}
  \right\rangle =
   \frac{\delta \mathcal L_g}{\delta g^{\mu \nu}}
  \left\langle \widehat{g}^{\mu \nu} \right\rangle =
\sqrt{-g} \, G_{\mu \nu} \left\langle \widehat{g}^{\mu \nu} \right\rangle.
\label{2-45}
\end{equation}

Now we have to make some physically reasonable
assumptions about the expectation value  $\left\langle \widehat{g}^{\mu \nu} \right\rangle$. We assume the following:
\begin{itemize}
  \item The expectation value
  $\left\langle \widehat{g}^{\mu \nu} \right\rangle$ is non-zero:
  \begin{equation}
	 \left\langle \widehat{g}_{\mu \nu} \right\rangle \neq 0~.
  \label{2-50}
  \end{equation}
  \item As a starting approximation,
  $\left\langle \widehat{g}_{\mu \nu} \right\rangle$ can be expressed in terms of some tensor of rank two
  constructed from the metric $g_{\mu \nu}$:
  \begin{equation}
	 \left\langle \widehat{g}_{\mu \nu} \right\rangle =
  K_{\mu \nu}.
  \label{2-60}
  \end{equation}
  \item The tensor $K_{\mu \nu}$ should have the same symmetry as the metric $g_{\mu \nu}$:
  \begin{equation}
	 K_{\mu \nu} = K_{\nu \mu}.
  \label{2-70}
  \end{equation}
  \item Taking into account the symmetry properties mentioned above, on can
  see that there exist the following possibilities of choosing
  the tensor $K_{\mu \nu}$:
	\begin{itemize}
	\item $K_{\mu \nu}$  is proportional to the metric tensor:
	\begin{equation}
		K_{\mu \nu} \propto g_{\mu \nu}.
	\label{2-80}
	\end{equation}	
	\item $K_{\mu \nu}$  is proportional to the Ricci tensor:
	\begin{equation}
		K_{\mu \nu} \propto \frac{R_{\mu \nu}}{R}.
	\label{2-90}
	\end{equation}		
	\end{itemize}
	\item The proportionality coefficient in the expressions \eqref{2-80} and
  \eqref{2-90} should be some invariant. Consequently, it must have the form
  $F(R, R_{\mu \nu} R^{\mu \nu}, \cdots)$. This yields
  \begin{itemize}
    \item
    \begin{equation}
	   \left\langle \widehat{g}_{\mu \nu} \right\rangle =
      K_{\mu \nu} =
      F(R, R_{\mu \nu} R^{\mu \nu}, \cdots)g_{\mu \nu};
    \label{2-93}
    \end{equation}
    \item
    \begin{equation}
	   \left\langle \widehat{g}_{\mu \nu} \right\rangle =
      K_{\mu \nu} =
      F(R, R_{\mu \nu} R^{\mu \nu}, \cdots)
      \frac{R_{\mu \nu}}{R};
    \label{2-96}
    \end{equation}
    \item or a linear combination of \eqref{2-93} and \eqref{2-96}. For example, it could be the Einstein or Schouten tensors.
  \end{itemize}
  The coefficient $F$ should be very small as $g_{\mu \nu} \rightarrow \eta_{\mu \nu}$, where $\eta_{\mu \nu}$ is
the Minkowski metric.	
\end{itemize}

Thus the Lagrangian \eqref{2-40} with the aforementioned
assumptions about the expectation value of the quantum part $\widehat g_{\mu \nu}$ of the metric
$\mathcal G_{\mu \nu}$ has the form
\begin{equation}
  \left\langle 	
		\mathcal L_{\mathcal G}(g + \widehat g)
	\right\rangle \approx - \frac{c^2}{2 \varkappa} \sqrt{-g}
  \left(
    R + G_{\mu \nu} K^{\mu \nu}
  \right).
\label{2-95}
\end{equation}
Hence we see that the quantum corrections coming from a non-zero expectation value of the nonperturbatively quantized metric give rise to modified gravity theories.

Let us now perform similar calculations for the matter Lagrangian, applying the decomposition \eqref{2-10}.
For simplicity, we consider a scalar field for which the Lagrange density is
\begin{equation}
	\mathcal L_m^{\mathcal G} =
	\sqrt{-\mathcal G}	\left[\frac{1}{2}\nabla^\mu \phi \nabla_\mu \phi - 	V(\phi)\right]\equiv
\sqrt{-\mathcal G}\mathcal L_m(g),
\label{2-100}
\end{equation}
with the classical Lagrangian
$$
\mathcal L_m(g)=\frac{1}{2}\nabla^\mu \phi \nabla_\mu \phi - 	V(\phi).
$$
To begin with,  we expand the quantum Lagrange density $\mathcal L_m^{\mathcal G}$ from Eq.~\eqref{2-100} as follows
\begin{equation}
	\mathcal L_m^{\mathcal G}(g + \widehat{g}) \approx \sqrt{-g} \mathcal L_m(g) +
  \frac{\delta \sqrt{-g}\mathcal L_m}{\delta g^{\mu \nu}}
  \widehat{g}^{\mu \nu}.
\label{2-110}
\end{equation}
The first variation of $L_m$ is well known:
\begin{equation}
	\frac{\delta \sqrt{-g}\mathcal L_m}{\delta g^{\mu \nu}} =
  \frac{\sqrt{-g}}{2}
	\left[		
		\nabla_\mu \phi \nabla_\nu \phi -
		g_{\mu \nu} \mathcal L_m
	\right] =
	\frac{\sqrt{-g}}{2} T_{\mu \nu},
\label{2-120}
\end{equation}
where $T_{\mu \nu}$ is the energy-momentum tensor. Consequently, we have
\begin{equation}
	\left\langle \mathcal L_m^{\mathcal G}(g + \widehat{g}) \right\rangle =
  \sqrt{-g}\left[\frac{1}{2}\nabla^\mu \phi \nabla_\mu \phi - V(\phi)
 + \frac{1}{2}T_{\mu \nu} K^{\mu \nu}\right].
\label{2-130}
\end{equation}
Thus we have obtained a scalar field nonminimally coupled to gravity.

Let us now consider a few cases with various tensors $K_{\mu \nu}$.

\subsection{The case $K_{\mu \nu} = g_{\mu \nu} \alpha$}

Substituting \eqref{2-93} into \eqref{2-95} and \eqref{2-130}, we obtain
\begin{eqnarray}
	\left\langle 	
		\mathcal L_{\mathcal G}(g + \widehat g)
	\right\rangle &\approx& - \frac{c^2}{2 \varkappa}
  \left(
    1 - \alpha
  \right) R \sqrt{-g},
\label{2-a-10}\\
  \left\langle \mathcal L_m^{\mathcal G}(g + \widehat{g}) \right\rangle
  &\approx& \sqrt{-g}\left[\mathcal L_m +
   \; \alpha T^\mu_\mu\right].
\label{2-a-20}
\end{eqnarray}
It is seen that Eq.~\eqref{2-a-10} assumes the redefinition of the gravitational constant $\varkappa$.

\subsection{The case $K_{\mu \nu} = - \Lambda \frac{g_{\mu \nu}}{R}$}

Again, substituting \eqref{2-93} in \eqref{2-95} and \eqref{2-130}, we obtain
\begin{eqnarray}
	\left\langle 	
		\mathcal L_{\mathcal G}(g + \widehat g)
	\right\rangle &\approx& - \frac{c^2}{2 \varkappa}
  \left(
    R + \Lambda
  \right) \sqrt{-g},
\label{2-a-30}\\
  \left\langle \mathcal L_m^{\mathcal G}(g + \widehat{g}) \right\rangle
  &\approx& \sqrt{-g}\left[\mathcal L_m -
   \; \frac{\Lambda}{R} T_\mu^\mu\right]~.
\label{2-a-40}
\end{eqnarray}
Hence we see from \eqref{2-a-30} that we have obtained Einstein gravity with a cosmological $\Lambda$-term
and matter nonminimally coupled to gravity [see Eq.~\eqref{2-a-40}].
If $\Lambda/R \ll 1$ then the coupling between matter and gravity becomes minimal, and we have
\begin{equation}
	\left\langle 	
		\mathcal L(g + \widehat g)
	\right\rangle =
  \left\langle 	
		\mathcal L_{\mathcal G}(g + \widehat g) + \mathcal L_m^{\mathcal G}(g + \widehat{g})
	\right\rangle \approx
\sqrt{-g}\left[  - \frac{c^2}{2 \varkappa}
  \left(
    R + \Lambda
  \right) + \frac{1}{2}	\nabla^\mu \phi \nabla_\mu \phi - 	V(\phi)\right].
\label{2-a-50}
\end{equation}

\subsection{The case $K_{\mu \nu} = \alpha R_{\mu \nu}$}

Substituting \eqref{2-96} into \eqref{2-95} and \eqref{2-130}, we obtain
\begin{eqnarray}
	\left\langle 	
		\mathcal L_{\mathcal G}(g + \widehat g)
	\right\rangle &\approx& - \frac{c^2}{2 \varkappa}
  \left(
    R - \frac{\alpha}{2} R^2 +
    \alpha R_{\mu \nu} R^{\mu \nu}
  \right) \sqrt{-g}~,
\label{2-a-60}\\
  \left\langle \mathcal L_m^{\mathcal G}(g + \widehat{g}) \right\rangle
  &\approx& \sqrt{-g}\left(\mathcal L_m +
  \frac{\alpha}{2}  \; T_{\mu \nu} R^{\mu \nu}\right),
\label{2-a-70}
\end{eqnarray}
where the dimensions of $\alpha$ is $\text{cm}^2$.

\subsection{The case $K_{\mu \nu} = \alpha G_{\mu \nu}$}

Representing the linear combination of \eqref{2-93} and \eqref{2-96} as the Einstein tensor
and substituting it into \eqref{2-95} and \eqref{2-130}, we obtain
\begin{eqnarray}
	\left\langle 	
		\mathcal L_{\mathcal G}(g + \widehat g)
	\right\rangle &\approx& - \frac{c^2}{2 \varkappa}
  \left(
    R + \alpha G_{\mu \nu} G^{\mu \nu}
  \right) \sqrt{-g},
\label{2-a-80}\\
  \left\langle \mathcal L_m^{\mathcal G}(g + \widehat{g}) \right\rangle
  &\approx& \sqrt{-g}\left(\mathcal L_m +
  \frac{\alpha}{2}  \; G^{\mu \nu} T_{\mu \nu}\right).
\label{2-a-90}
\end{eqnarray}

\section{Discussion and conclusions}

Thus we have shown that if a metric in quantum gravity can be represented as a sum of classical and quantum parts then such a gravitating physical system looks approximately  as being described by
modified gravity, in which a nonminimal interaction between matter and gravity is present.
The modification of gravity occurs as a consequence of the presence of a non-zero
 expectation value of the quantum part of the metric. 

\textcolor{blue}{In particular, proceeding along these lines, by a specific choice for
the expectation value of the quantum metric,
we can obtain a $\Lambda$-term
as a consequence of the fluctuating part of the metric. 
Such a model then explains qualitatively the smallness of the cosmological term:}
it is just the expectation value of quantum fluctuations of the metric. At the present epoch these fluctuations
should be very small. Following this way, we have also shown that quantum fluctuations lead to a nonminimal interaction between gravity and matter. 

\textcolor{blue}{We note here that attempts to obtain the cosmological constant 
from quantum effects are not new. In curved spacetime the regularization of quantum fields 
gives rise to a one-loop effective Lagrangian for the gravitational field, 
which (for example, for an FRW universe) becomes:
$
\mathcal L_{eff} = \Lambda_\infty +
R/16 \pi G_{\infty} +
\alpha_{\infty} R^2 +
\beta_{\infty} R_{\mu \nu} R^{\mu \nu}
$ (for details, see e.g.~Refs.~\cite{Sahni:1999gb,Birrell}). 
Unfortunately, such calculations do not give good values for the induced cosmological constant. 
For example, for the best QCD estimation the vacuum energy density is 
$\rho_{QCD} \sim 10^{-3}$GeV$^4$, 
which is considerably larger than 
the observed value $\rho_{vac} \sim 10^{-47}$GeV$^4$. 
In contrast, in our approach employed here quantum corrections arise from the
nonperturbative quantization of the metric, 
and not from the perturbative quantization of various types of fields 
(scalar, electromagnetic, QCD).
}

It must be emphasized that the model under consideration assumes that for each quantum state $\left|Q\right\rangle$
there exists a unique function $F(R,R_{\mu \nu}R^{\mu \nu}, \cdots)$.
Thus one can say that $F$-gravities are dynamical ones in the sense that for different quantum states we have different functions $F$.

\section*{Acknowledgements}

This work was supported by the Volkswagen Stiftung.
V.D. and V.F. acknowledge support from a grant
in fundamental research in natural sciences
by the Ministry of Education and Science of Kazakhstan.
B. K. and J. K. acknowledge support by the DFG Research Training Group 1620 ``Models of Gravity''.

\end{document}